# Towards more realistic dynamical models for DNA secondary structure

Sahin BUYUKDAGLI, Michaël SANREY and Marc JOYEUX[(#)]

*Laboratoire de Spectrométrie Physique (CNRS UMR 5588),*

*Université Joseph Fourier - Grenoble 1,*

*BP 87, 38402 St Martin d'Hères, FRANCE*

[(#)] email : Marc.JOYEUX@ujf-grenoble.fr

**Abstract :** We propose a dynamical model for the secondary structure of DNA, which is based on the finite stacking enthalpies used in thermodynamics calculations. In this model, the two strands can separate and the bases are allowed to rotate perpendicular to the sequence axis. We show, through molecular dynamics simulations, that the model has the correct behaviour at the denaturation transition.

As is well-known, DNA is a set of two entangled polymers. Each monomer, called nucleotide, is composed of three elements, namely a phosphate group, a five-atomic sugar ring, and a base (cytosine (C) and thymine (T) are monocyclic, guanine (G) and adenine (A) are bicyclic). The backbone of each strand consists of alternating phosphate and sugar groups. Each base is attached to a sugar group with its ring(s) roughly perpendicular to the sequence axis. The collection of letters A, T, G and C, which describes a specific DNA sequence and encodes the genetic information of the molecule, is known as the "primary structure", or "genome", of the DNA. At body temperature, DNA is usually observed in the double-stranded structure, which results from the association of two polymers of single-stranded DNA. Watson and Crick showed that this association is due to the hydrogen bonds, which are formed selectively between A and T and between G and C. The base-pairing of the two polymers is commonly referred to as the "secondary structure" of DNA. In addition, DNA sequences also have well-defined higher order conformations, like the B-, A- and Z- double helix forms, known as the "tertiary structure", which are however disregarded in this Letter.

The two strands of DNA separate upon heating. This phenomenon, called denaturation or melting [1,2], occurs through series of steps, which can be monitored by UV absorption spectroscopy [3]. At each step, large portions of the inhomogeneous DNA sequence separate over very short temperature intervals. The whole denaturation process therefore looks like a chain of sharp, first order-like phase transitions. At the scale of the bases, the simplest description of melting assumes, like the Ising model, that a pair is either open or closed and that its evolution can be depicted by a two-state (0 or 1) variable. Starting with the work of Poland and Sheraga [4], a large number of models were based on this assumption. Some of them are now commonly and successfully used to compute denaturation curves of homogeneous and inhomogeneous DNA, which are in good agreement with experiment [5,6].

While statistical models were specifically derived to investigate the denaturation of DNA, dynamical models based on a Hamiltonian function of continuous variables can be expected to describe the whole dynamics of DNA, which extends from small-amplitude oscillations at low temperatures to large-amplitude motions close to denaturation. The more ancient models, which were proposed to study soliton or solitary excitations in DNA (see Ref. [7] and references therein), can however not reproduce denaturation, because the only degree of freedom of each nucleotide is the rotation angle of the base around the strand axis. This is no longer the case for the model proposed by Prohofsky and co-workers (see Ref. [8] and references therein), which consider instead that the principal source of non-linearity in DNA is the hydrogen bond between paired bases (taken as a Morse potential), and that the important degree of freedom is the corresponding stretching coordinate. Dauxois, Peyrard and Bishop (DPB) later replaced, in the model of Prohofsky and co-workers, the harmonic stacking interaction between two successive bases by an anharmonic one and showed that this leads to denaturation curves in better agreement with experiment [9]. The DPB model was subsequently used to unravel several aspects of homogeneous DNA dynamics (for a recent review see Ref. [10]).

The DPB model does however not take into account the fact that stacking interactions necessarily vanish when two successive bases have slid on each other far enough for their rings not to be superposed any longer. Therefore, we recently proposed another expression for the nearest-neighbour staking interaction, which is based on the finite propagation enthalpies used in thermodynamic calculations [6], and showed that the finiteness of the nearest-neighbour potential is sufficient to induce a first-order phase transition at the melting temperature [11]. Moreover, we showed that this model reproduces correctly the multi-step denaturation process, which is experimentally observed for inhomogeneous DNA sequences of length 1000-10000 bp [11]. This had never been achieved before with a dynamical model and indicates that site-specific stacking interactions are a fundamental ingredient for the correct dynamical description of inhomogeneous DNA sequences.

Despite these encouraging results, the model of Ref. [11] is still pretty rough. To our mind, its two principal defects are both related to the low number of degrees of freedom which are taken into account. Indeed, a single particle describes a complete monomer, that is, a particular base and the attached sugar and phosphate groups, while the bases and the backbones necessarily have quite different dynamics. Moreover, the model assumes that the unstacking of two successive bases and the separation of the two strands are both governed by the same (essentially stretching) coordinate, whereas one might reasonably expect, in agreement with the older models [7], that rotation of the bases perpendicular to the sequence axis plays an important role in the unstacking process. The purpose of this Letter is to propose an advance of the model of Ref. [11], which accommodates more properly these two points, and to show, through molecular dynamics simulations, that the improved model still displays the correct melting behaviour.

The improved model is schematized in Fig. 1. Each backbone is represented by an array of point masses $m$ tied together by springs. A point mass $m$ thus describes a phosphate group plus a sugar group - but not a base. In order to limit the number of degrees of freedom, the point masses $m$ are regularly spaced along the sequence axis $z$ and can move only along the axis $x$, which connects the two backbones. $x_n$ and $X_n$ indicate the respective locations along the $x$ axis of the $n^{th}$ point mass of each backbone. The bases are represented by two other sets of point masses $M$, each point mass $M$ being rigidly attached to a phosphate/sugar point mass $m$. While the distance $R$ between a base and the phosphate/sugar group to which it is attached is fixed, the base is allowed to rotate in the plane $(x, y)$ perpendicular to the sequence axis $z$. The position of each base thus depends on the torsion angle $\gamma_n$ (respectively $\Gamma_n$) in addition to the translation coordinate $x_n$ (respectively $X_n$). Two neighbouring bases belonging to the same strand interact together through a finite nearest-neighbour stacking potential similar to that of Ref. [11], while two paired bases located on opposite strands interact through the usual Morse potential. This is however not sufficient. At this point, there is indeed nothing that prevents the double-stranded DNA sequence from folding. A term, which maintains the four point masses at position $n$ more or less aligned up to the dissociation threshold of the Morse potential, must therefore be introduced in order to prevent this unphysical behaviour. This term mimics the action of the two or three hydrogen bonds, which connect each pair of bases along different directions and efficiently prevent folding. Denoting by $u_n = x_n + R\cos\gamma_n$, $v_n = R\sin\gamma_n$, $U_n = X_n + R\cos\Gamma_n$ and $V_n = R\sin\Gamma_n$ the coordinates of the point masses $M$ in the $(x,y)$ plane, and by $d_n = \sqrt{(U_n - u_n)^2 + (V_n - v_n)^2}$ the distance between two paired bases, the Hamiltonian of the model we propose is thus

$$H = T + V$$

$$T = \frac{1}{2}(m+M)\sum_n \left(\dot{x}_n^2 + \dot{X}_n^2\right) + \frac{1}{2}\frac{mMR^2}{m+M}\sum_n \left(\dot{\gamma}_n^2 + \dot{\Gamma}_n^2\right)$$

$$V = D\sum_n \{1 - \exp[a(d_0 - d_n)]\}^2$$

$$+ K_b \sum_n (2 - \cos\gamma_n + \cos\Gamma_n)\exp[a(d_0 - d_n)] \qquad (1)$$

$$+ \frac{1}{2}\sum_n \Delta H^{(n)}\left\{1 - \exp\left[-b(u_n - u_{n+1})^2 - b(v_n - v_{n+1})^2\right]\right\}$$

$$+ \frac{1}{2}\sum_n \Delta H^{(n)}\left\{1 - \exp\left[-b(U_n - U_{n+1})^2 - b(V_n - V_{n+1})^2\right]\right\}$$

$$+ K_s \sum_n \left[(x_n - x_{n+1})^2 + (X_n - X_{n+1})^2\right] .$$

A simplified form has been assumed for kinetic energy $T$, so that a second-order Brünger-Brooks-Karplus integrator can be used in molecular dynamics simulations (see below). The first term in the right-hand side of the potential energy $V$ is the Morse potential, which models the hydrogen bonds that connect a base pair. The second term prevents folding of the two strands, as discussed just above. The third and fourth terms describe the nearest-neighbour stacking interactions between successive bases belonging to the same strand. Finally, the last term represents the elastic energy of the two phosphate/sugar backbones.

As in Ref. [11], the finite stacking enthalpies $\Delta H^{(n)}$, which range from 0.347 to 0.465 eV, are borrowed from thermodynamics studies, more precisely from Table 1 of Ref. [6]. The values of most of the other parameters are either derived from known chemical data : $m$=180 amu, $M$=130 amu, $d_0$=4 Å, $R$=7 Å and $b$=0.05 Å$^{-2}$, or transferred from similar models : $D$=0.05 eV and $a$=3.5 Å$^{-1}$ [9-11]. On the other hand, $K_b$=0.5 eV and $K_s$=0.2 eV Å$^{-2}$ were roughly adjusted by hand in order that the denaturation curves obtained from the model agree with experiment. In order to indicate to what extent the results presented below are sensitive to the values of these parameters, let us mention that assuming $K_b$=0.1 eV (resp. $K_b$=1.0 eV), instead of $K_b$=0.5 eV, decreases (resp. increases) the melting temperature of all sequences by about 40 K. $K_b$ should consequently be taken in the range 0.1-0.6 eV. Similarly, assuming $K_s$=0.1 eV Å$^{-2}$ instead of $K_s$=0.2 eV Å$^{-2}$ decreases the difference between the melting temperatures of homogeneous AT and GC sequences from 40 K down to 25 K (see below). $K_s$ is therefore rather rigidly determined for given values of the parameters of the Morse oscillator.

The denaturation dynamics of the model of Eq. (1) was studied numerically by integrating Langevin's equations of motion with a second-order Brünger-Brooks-Karplus integrator [12]. This recurrence relation connects the values taken by each coordinate $q_n$ ($q_n$=$x_n$, $X_n$, $\gamma_n$ or $\Gamma_n$) at successive time intervals $(k-1)\tau$, $k\tau$, and $(k+1)\tau$ according to

$$\frac{\mu}{\tau^2}\left(q_n^{k+1} - 2q_n^k + q_n^{k-1}\right) = -\left.\frac{\partial V}{\partial q_n}\right|_{k\tau} - \frac{\mu\gamma}{2\tau}\left(q_n^{k+1} - q_n^{k-1}\right) + w(n,k)\sqrt{\frac{2\mu\gamma k_B T}{\tau}} , \qquad (2)$$

where γ is the dissipation coefficient, $w(n,k)$ a normally distributed random function with zero mean and unit variance, and μ is equal to $m+M$ for a stretching coordinate ($q_n = x_n$ or $X_n$) and to $mMR^2/(m+M)$ for a bending one ($q_n = \gamma_n$ or $\Gamma_n$). The second and third term in the right-hand side of Eq. (2) model the effects of the solvent on the DNA sequence. Calculations were performed with a time step τ=10 fs, that is, about one hundredth of the period of the Morse oscillator at low temperature. We checked on a few points that reduction of τ by a factor ten does not change the results presented below. Assumed value for the dissipation coefficient is γ=5 ns$^{-1}$. Although the results presented below may well depend on this value - for example, the sequence cannot be heated to sufficiently high temperature if γ is too large -, we did not investigate in detail the influence of γ on the present model.

The profiles at various temperatures for a 1793 bp ATATAT... sequence and the 1793 bp inhomogeneous sequence with NCBI entry code NM_001101 are shown in Figs. 2 and 3, respectively. Each profile displays, as a function of the base position $n$, the average of $d_n - d_0$ over 0.15 μs integration times. It is seen in Fig. 2 that, as expected, the profiles of the ATATAT... sequence are essentially flat, except for the base pairs located close to the ends of the sequence, which open at slightly lower temperatures. Because of the sharpness of the transition, $\langle d_n \rangle - d_0$ increases much more rapidly with temperature close to the denaturation threshold than at lower temperatures.

Comparison of the two plots of Fig. 3 indicates that the profiles of inhomogeneous sequences are not flat but reflect instead the local AT percentage : remember that the higher the content of A and T bases in two successive base pairs, the lower the stacking enthalpy $\Delta H^{(n)}$ [6]. More precisely, the bottom plot of Fig. 3 shows the AT percentage averaged over 40 consecutive base pairs of the sequence. This percentage globally increases from $n$=1 to $n$=1793, with three local maxima at $n$≈1300, $n$≈1450 and $n$≈1600. Examination of the profiles in the top plot shows that melting correspondingly starts in narrow regions centred around these three peaks, while the sequence abruptly melts for all $n \geq 1000$ at slightly higher temperatures. One then observes a plateau of about 3-4 K before the lower end of the sequence sharply melts. Note that these domains are in excellent agreement with those obtained from statistical mechanics calculations (see Fig. 2 of ref. [13]).

Current experiments are not able to provide as detailed information as the profiles of Figs. 2 and 3. Nonetheless, extinction curves obtained from UV absorption spectroscopy around 270 nm can be compared directly with the calculated curves for the fraction of open base pairs (see Ref. [6] and references therein). While the distinction between closed and open base pairs lies at the heart of statistical models, these concepts need to be defined somewhat arbitrarily for dynamical models. After some trials, we found that a reasonable and convenient choice consists in considering that the base pair at site $n$ is closed as long as $\langle d_n \rangle - d_0 \leq 9$ Å and open when $\langle d_n \rangle - d_0 > 9$ Å. The threshold value of 9 Å corresponds to the situation where the four point masses describing site $n$ have remained aligned but the distance between the two backbones has increased by 50% (i.e. from the equilibrium value of 18 Å up to 27 Å), or, alternatively, to the situation where the distance between the two backbones has remained fixed to 18 Å but the paired bases have rotated by 45° in opposite directions or 70° in the same direction. The threshold value of 9 Å is indicated by an horizontal dash-dotted line in Figs. 2 and 3.

The denaturation curves obtained with this criterion are shown as empty circles connected by solid lines in Fig. 4 for the two 1793 bp sequences of Figs 2 and 3 (i.e. an ATATAT... sequence and the inhomogeneous sequence with NCBI entry code NM_001101), as well as a 1793 bp GCGCGC... sequence. As for Figs. 2 and 3, each point was obtained by averaging $d_n - d_0$ over 0.15 μs integration times. This figure shows that the model of Eq. (1)

successfully reproduces the four principal properties of experimental denaturation curves : (i) melting of homogeneous sequences occurs in very narrow temperature windows, (ii) the melting temperatures of pure AT and pure GC sequences are separated by a gap of about 40 K, (iii) the melting temperature of inhomogeneous sequences is proportional to the AT (or GC) percentage, (iv) denaturation of inhomogeneous sequences occurs through a series of jumps. This confirms that use of sequence-specific stacking interactions is of primary importance for the correct description of the dynamics of inhomogeneous DNA sequences.

In order to show that the 9 Å threshold is by no means critical, we also plotted in Fig. 4 the denaturation curves obtained with 6 Å (× symbols) and 12 Å (crosses) thresholds. It is seen that these values essentially shift the denaturation curves to, respectively, lower and higher temperatures by a couple of Kelvins. Moreover, the jumps are slightly less marked with the 12 Å criterion.

Before concluding, let us finally note that calculations indicate that there is a slight interval between the temperature where all the base pairs are considered to be open and the temperature where the two strands separate effectively. Indeed, in the former case, the *time average* of $d_n - d_0$ must be larger than the 9 Å threshold for all *n*, while, in the later case, all $d_n - d_0$ must be larger than 9 Å *simultaneously*. The difference between the two temperatures is pretty small (1-2 K) for the inhomogeneous and GCGCGC... sequences, but reaches 5-6 K for the ATATAT... sequence (see Fig. 2). The last point drawn at the high-temperature end of each plot of Fig. 4 corresponds precisely to the temperature where the two strands first separate.

In conclusion, we have proposed a dynamical model for the secondary structure of DNA, which is based on the finite stacking enthalpies used in thermodynamics calculations and has the correct behaviour at the denaturation transition. This model is an advance of our previous work [11], because the backbones are clearly separated from the bases, which are allowed to rotate perpendicular to the sequence axis. Of course, this improvement requires that a larger number of degrees of freedom be considered, that is, 4$N$ for a sequence of length $N$. It is now tempting to include the tertiary structure of DNA in the model and/or to investigate to what extent this kind of model can be used to study the interaction of DNA sequences with other molecules with specific biological role.

**FIGURE CAPTIONS**

**Figure 1** (color online) : Schematic view of the proposed model for the secondary structure of DNA.

**Figure 2** (color online) : Plot, at increasing temperatures $T$, of $\langle d_n \rangle - d_0$ as a function of the site number $n$ for a 1793 bp ATATAT... sequence.

**Figure 3** (color online) : (Top) Plot, at increasing temperatures $T$, of $\langle d_n \rangle - d_0$ as a function of the site number $n$ for the 1793 bp sequence with NCBI entry code NM_001101. (Bottom) Plot, as a function of $n$, of the AT percentage averaged over 40 consecutive base pairs.

**Figure 4** (color online): Plot of the fraction of open base pairs (in %) as a function of temperature $T$ for a 1793 bp ATATAT... sequence, the 1793 bp inhomogeneous sequence with entry code NM_001101, and a 1793 bp GCGCGC... sequence. For each sequence, we show the curves obtained with three different criteria for open base pairs, namely $\langle d_n \rangle - d_0 > 6$ Å (× symbols), $\langle d_n \rangle - d_0 > 9$ Å (empty circles), and $\langle d_n \rangle - d_0 > 12$ Å (crosses) .

**Figure 1**

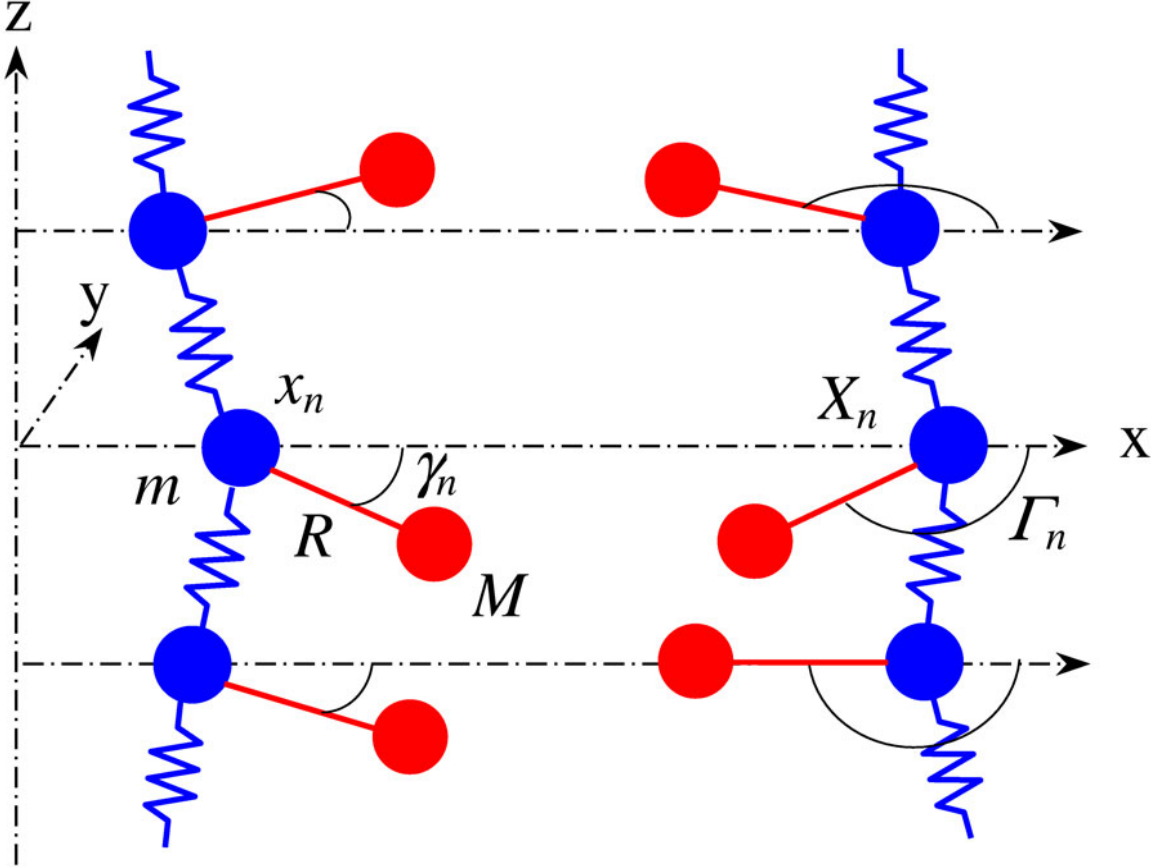

**Figure 2**

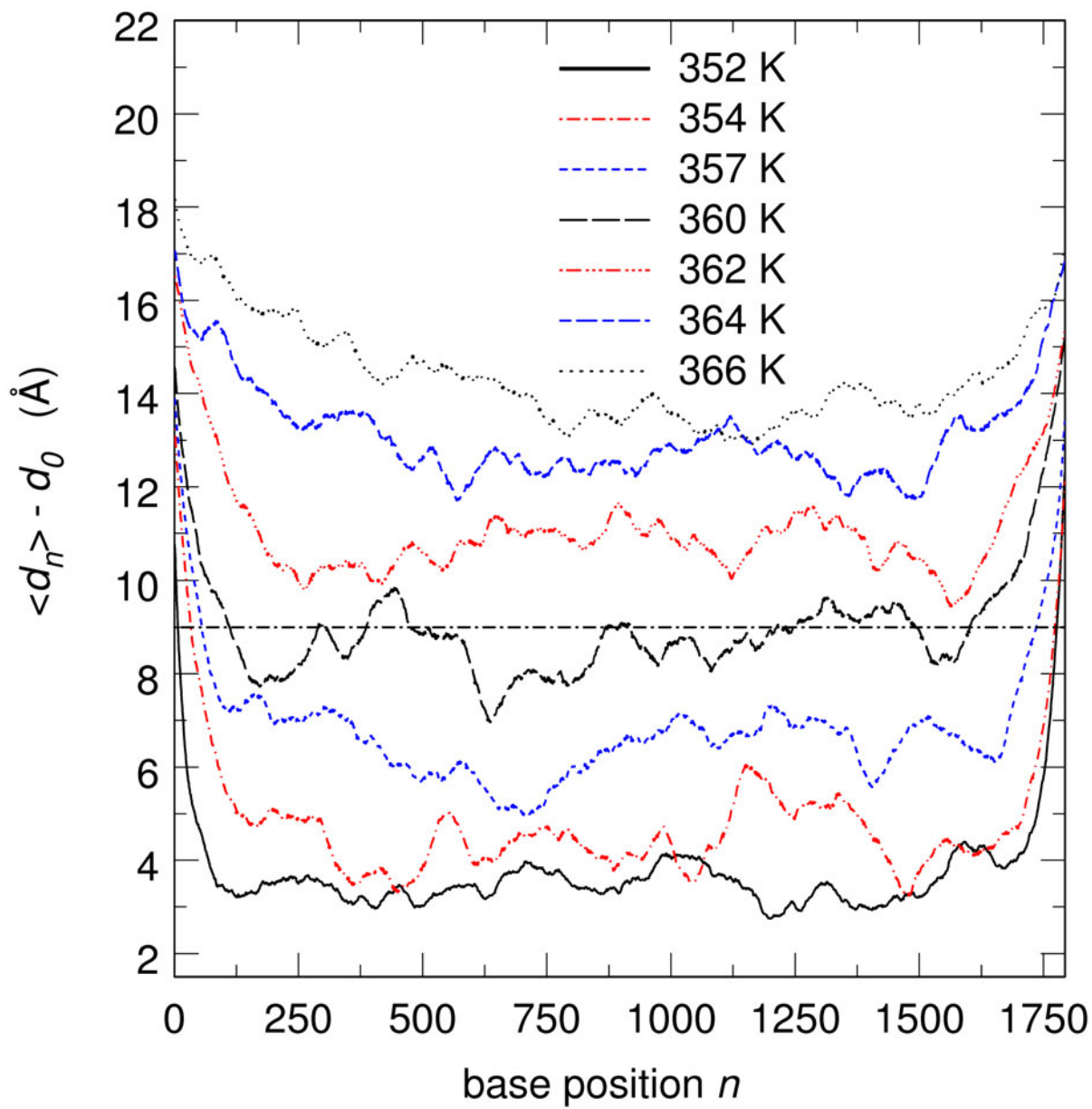

**Figure 3**

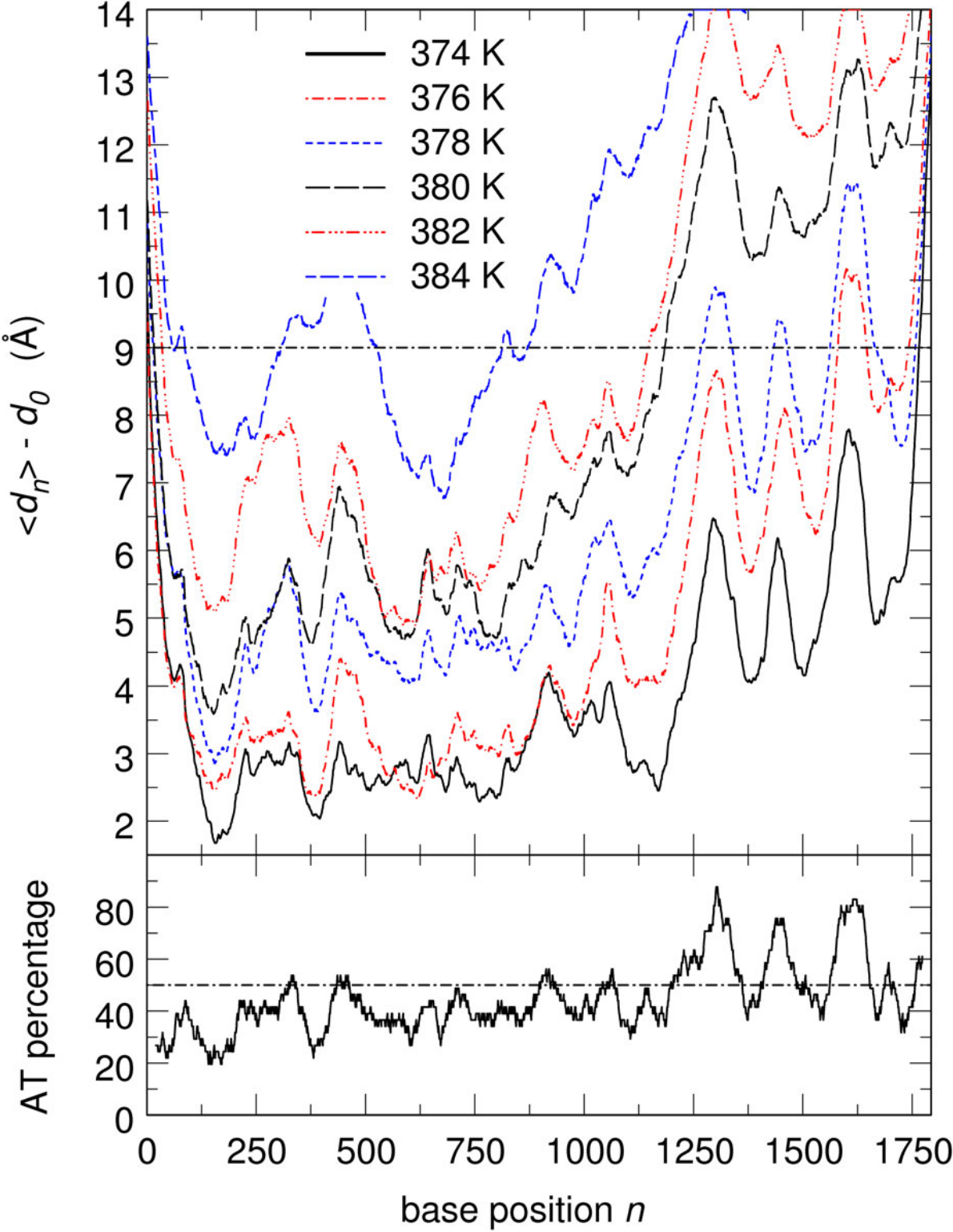

**Figure 4**

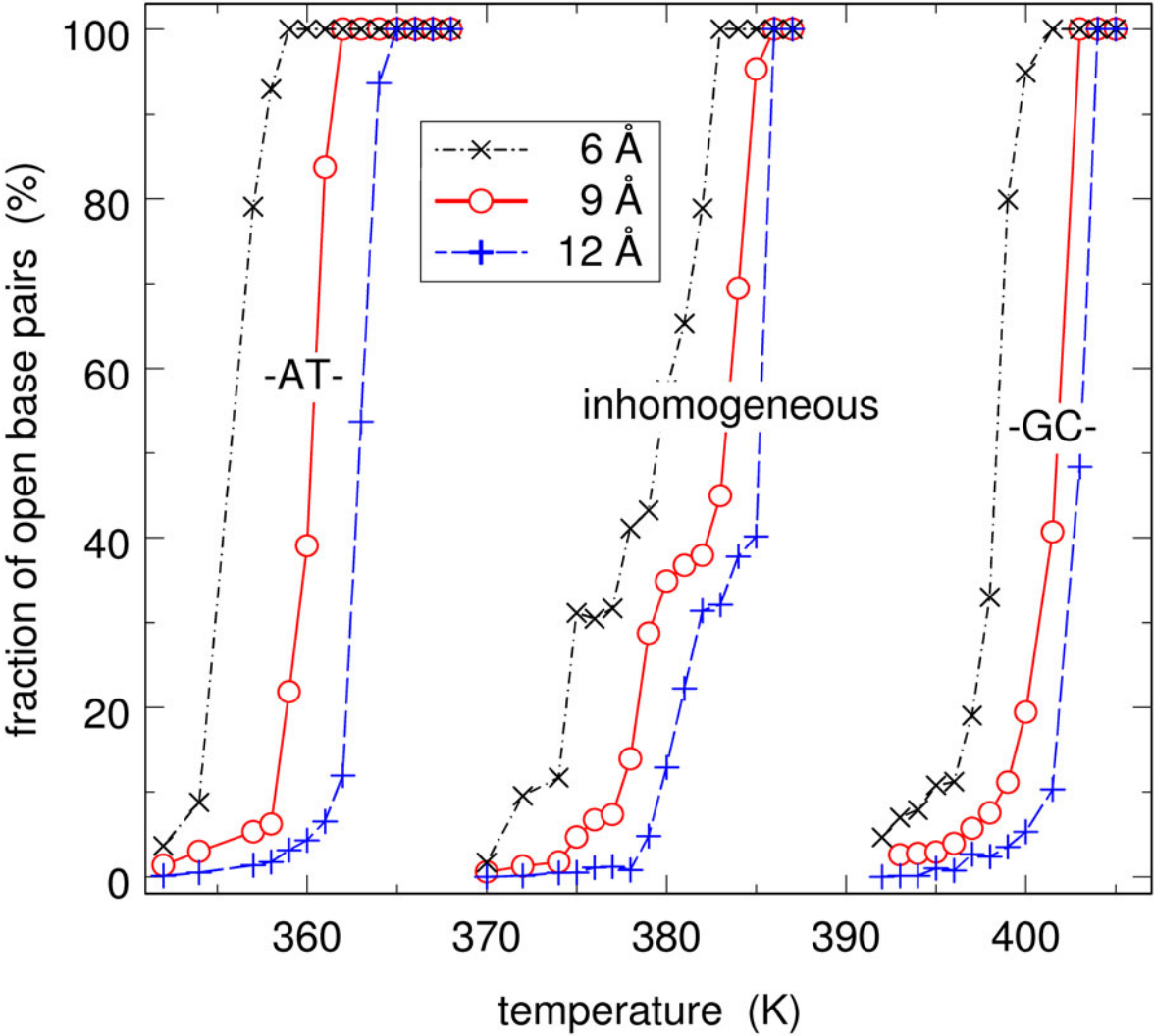